\documentclass{article}
\pdfoutput=1

\usepackage[preprint, nonatbib]{neurips_2019}

\usepackage[utf8]{inputenc} 
\usepackage[T1]{fontenc}    
\usepackage{hyperref}       
\usepackage{url}            
\usepackage{booktabs}       
\usepackage{amsfonts}       
\usepackage{nicefrac}       
\usepackage{microtype}      
\usepackage{times}
\usepackage{latexsym}
\usepackage{graphicx} 
\usepackage{url}
\usepackage{lipsum} 
\usepackage{multicol, blindtext}

\usepackage[font=small,skip=0pt]{caption}
\usepackage{amsmath}
\usepackage{stmaryrd}

\usepackage{amssymb}
\usepackage{hhline}
\usepackage{multirow}
\usepackage[usestackEOL]{stackengine}
\setstackgap{C}{\normalbaselineskip}
\title{Graph Star Net for Generalized Multi-Task Learning}

\author{%
  Lu Haonan \\
  AARC\\
  Huawei Technologies\\
  \texttt{luhaonan@huawei.com} \\
   \And
   Seth H. Huang
   \thanks{Corresponding Author.  Lu Haonan and Tian Ye are equal contributors}\\
   AARC \\
   Huawei Technologies \\
   \texttt{sethhuang@huawei.com} \\
   \AND
   Tian Ye \\
   AARC \\
   Huawei Technologies \\
   \texttt{tianye25@huawei.com} \\
   \And
   Guo Xiuyan \\
   AARC\\
   Huawei Technologies \\
  \texttt{guoxiuyan1@huawei.com} \\
}

\begin{document}
\maketitle
\begin{abstract}
In this work, we present graph star net (GraphStar), a novel and unified graph neural net architecture which utilizes message-passing relay and attention mechanism for multiple prediction tasks - node classification, graph classification and link prediction. GraphStar addresses many earlier challenges facing graph neural nets and achieves non-local representation without increasing the model depth or bearing heavy computational costs. We also propose a new method to tackle topic-specific sentiment analysis based on node classification and text classification as graph classification. Our work shows that “star nodes” can learn effective graph-data representation and improve on current methods for the three tasks. Specifically, for graph classification and link prediction, GraphStar outperforms the current state-of-the-art models by 2-5\% on several key benchmarks.
\end{abstract}

\section{Introduction}
Relational data, consisted of nodes, sub-graphs and edges, has a vast range of applications from sentiment analysis, text classification to protein and enzyme interactions. Graphs have implicit, complex topological structures and relationships, and graph models capable of internalizing these structures have achieved state-of-the-art performances for many task \cite{zhou2018graph}. For data representation, graphs are more universal than conventional pairwise relationships, and machine learning approaches for graph analysis mainly focus on three primary tasks: node classification, graph classification and link prediction \cite{monti2017geometric,perozzi2014deepwalk, tang2015line}. 

Graph data models a set of subjects/ entities (nodes) and their inter-relationships (edges). In a simplest classification setting, the model attempts to predict unlabeled nodes by the surrounding labeled nodes. The graph domain is also categorized into spectral and non-spectral approaches. Spectral approaches often work with methods to represent the graph. Non-spectral approaches often conduct local-convolutions on the graph in the neighborhood \cite{hamilton2017inductive,velivckovic2017graph}. In order to capture the global state, a local conventional model has to increase its depth, but this may encounter an over-smoothing issue \cite{zhou2018graph,gilmer2017neural,li2015gated}. On the other hand, \cite{wang2018non} tackles non-local representation to capture long-range dependencies, using fully-connected-nodes attention and aggregating all features of all positions as a weighted sum, resulting in a high computational complexity. Our model aims at capturing the global state without increasing the model depth and using fully-connected attention. The codes for essential experiments will be released at \url{ https://github.com/graph-star-team/graph_star}. This site complies fully with the anonymous, double-blinded policies.

Graph Attention Networks, instead of using convolutional neural networks (CNN) as in \cite{hamilton2017inductive}, proposes a graph attention layer, which compared to CNNs can specify different weights to different nodes within a neighborhood \cite{velivckovic2017graph}. They adopt the multi-head attention mechanism \cite{velivckovic2017graph,devlin2018bert}, allowing for predictions of new-node and previously unseen graphs, a transductive-to-inductive approach. 
However, it does not allow for deep-neighborhood representation, which lessens the model capability to understand context. \cite{wang2018non} addresses this issue with a self-attention approach to capture global state without increasing the model depth.

Star Transformer \cite{guo2019star} changes the original fully-connected attention structure to one with a virtual message-passing relay, reducing the number of connections from quadratic to linear \cite{guo2019star}. The design has the potential of capturing long-range dependency, aggregating both local and non-local compositionality \cite{guo2019star}. Our work has shown improvement upon previous results while simplifying the model structure and improving the computational efficiency. 

In this work, we introduce Graph Star Net (GraphStar), an inductive framework which extends the information boundary through relay nodes to aggregate global information. To our best knowledge, we have not seen another architecture which utilizes the virtual relay concept in graph modeling and aggregating local and non-local information. Our main contributions lie in addressing many previous limitations and unifying the approaches in multiple tasks for node classification, graph classification and link prediction simultaneously. We also proposed a new node-classification-based, topic-specific sentiment analysis method as well as graph-classification-based, text classification method. Compared with the previous node-classification-based transductive models such as \cite{DBLP:journals/corr/abs-1902-07153,yao2018graph}, our model can easily be used for unseen, new document predictions.

GraphStar utilizes the unique data representation via an information relay across the graph and trains the model to: (1) perform inductive tasks on previously unseen graph data; (2) aggregate both local and long-range information, making the model globally-aware, extracting high-level abstraction typically not represented in individual node features; (3) the relays serve as a hierarchical representation of the graphs and can be directly used for graph classification tasks.

\section{GraphStar Model Architecture}

In this section, we present the star relay function and the updating process and describe the theoretical and implementation advantages. Generally, our approaches will be divided into three steps: star initialization, updating of the real nodes and the updating of the stars. In each layer of the model, the second and third steps occur in a cyclical manner layer-by-layer in the updating process, and our goal is to train a multi-purpose model to represent local and global compositionality for different prediction tasks. Our model setup primarily follows the landmark works of \cite{velivckovic2017graph,wang2018non} as well as \cite{guo2019star}. 


\begin{figure}[htb]
      \center{\includegraphics[width=5in]
      {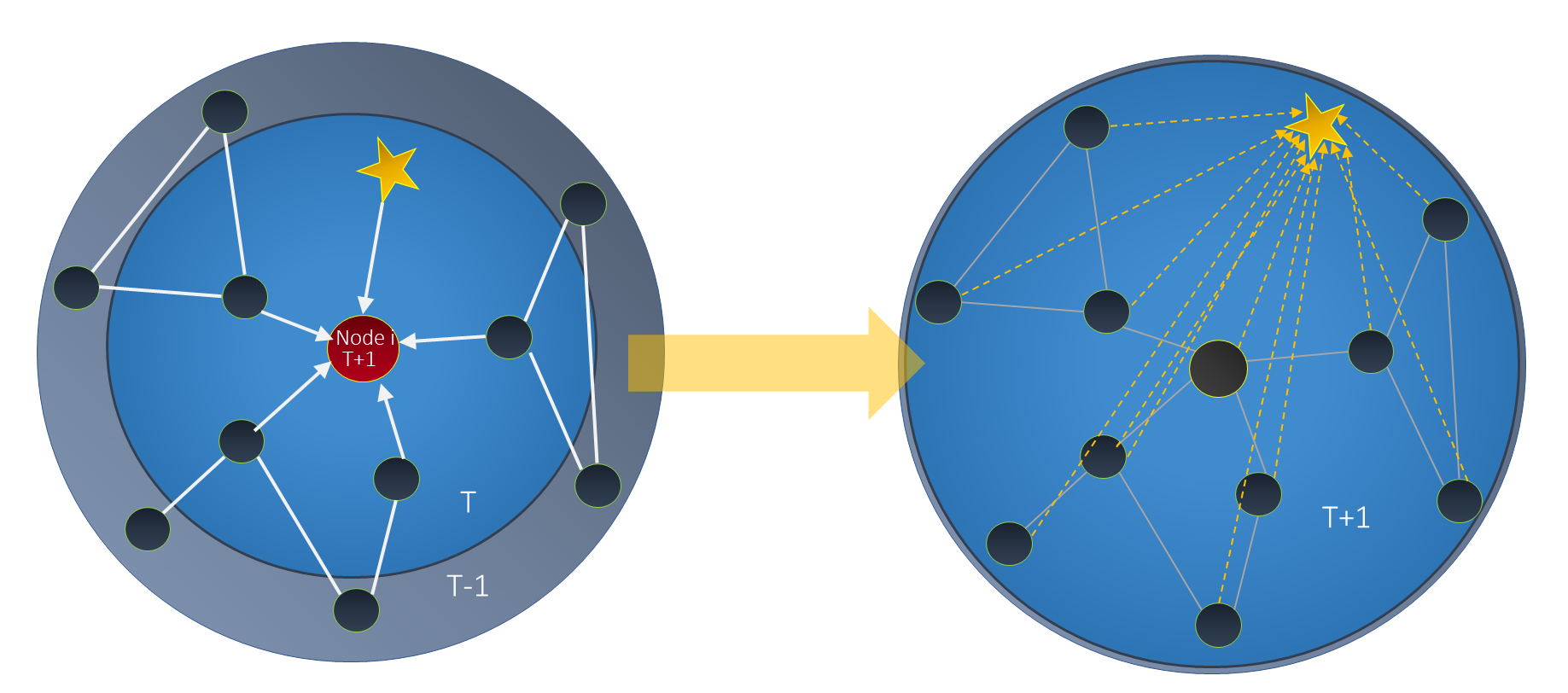}}
      \caption{\label{fig:Graph Star Architecture} Graph Star Architecture. Left: Corresponding to Step 2, local compositionality updates for real nodes. Node i uses the representations of its neighbors, including itself, and the star in the layer t to get its own representation in layer (t+1). Right: Corresponding to Step 3, star global compositionality updates. To update the star, the model uses the updated representations of all nodes in the layer (t+1) and its own representation in layer t to get star’s representation in (t+1). This star, connected to all nodes, is a natural hierarchical representation of the whole graph.}
\end{figure}

The input to our model is a graph \(G(A,F)\), where  \(A \epsilon\,R^{N_b \times N_b}\) is adjacency matrix and \(F\epsilon R^{n \times d}\) is the feature matrix, and each node has \(d\) features, and \(N_b\) is the number of nodes in a graph. We describe the graph with a set of learned features in each layer \(H^{t} = (\overrightarrow{h}_1,\overrightarrow{h}_2,...,\overrightarrow{h}_{N_b}),\overrightarrow{h}_i^t\epsilon\,\mathbb{R}^d \), and \(F = (\overrightarrow{f}_1,\overrightarrow{f}_2,...,\overrightarrow{f}_{N_b}),\overrightarrow{f}_i\epsilon\,\mathbb{R}^d \) is the initial representation of the node \(i\) and \(t\) being the layer number. 

\textbf{Step 1: Initial Representation of the Star}

Let \(\overrightarrow F_{mean} = mean(F^0), F\epsilon \mathbb{R}^{d}\). We have the initial star representation \(\overrightarrow S^{(0)} = \sigma(\sum\limits_{i\epsilon N_b} \alpha_{init,i}W_V^{init}\overrightarrow f_i)\), where
\begin{equation}
\alpha_{init,i} = \frac{\exp(<W_Q^{init}  \overrightarrow{F}_{mean}, W_K^{init} \overrightarrow{f}_i>)}{\sum\limits_{k\epsilon N_b}{\exp(<W_Q^{init}  \overrightarrow{F}_{mean}, W_K^{init} \overrightarrow{f}_k>)}}
\end{equation}

\(<\overrightarrow a,\overrightarrow b>\) is the dot product of \(\overrightarrow a,\overrightarrow b\), \(\sigma\) is the non-linear activation function and \(W_Q^{init},W_K^{init},W_V^{init}\) follow the standard Transformer setup \cite{vaswani2017attention}. 

\vspace{1mm}

\textbf{Step 2: Real Node Update}

Here we use multi-head attention to conduct real node update \cite{velivckovic2017graph}. Inspired by rGCN \cite{schlichtkrull2018modeling}, we treat node-to-neighborhood under relation \(r\epsilon R\), node-to-star and node-to-self as three types of relations, with the first one corresponding to the number of relations in the graph. The node importance under relation \(r\) is controlled by attention coefficients.
We define the self-self, self-star and self-neighbor under relation \(r\) as controlled by \(W_{V0}\), \(W_{VS}\) and \(W_{rV1}\), respectively. \textbf{The work presented in our paper is based on the special case \(R=1\), making it single-type relation link prediction}. The node update can be represented by

\begin{equation}
h_i^{(t+1)} = \bigparallel_m^{N_{Head}} \sigma(\sum_{r\epsilon R}\sum_{j\epsilon N_i^r} \alpha_{ijr}^m W_{rV1}^{m(t)} h_j^{(t)}+\alpha_{is,r=s}^m W_{VS}^{m(t)}S^t + \alpha_{i0,r=0}^m W_{V0}^{m(t)}h_i^t),
\end{equation}

where \(\bigparallel\) represents the concatenation of all multi-heads, \(N_i^r\) is the neighbors of node i under relation \(r\epsilon R\).
\(N_{head}\) is the number of heads in the multi-head attention setting. The multi-head outputs are then concatenated as the representation of \(t+1\) layer. Here we define

\[
\hat{\sum}^m = \sum_{r\epsilon R}\sum_{k\epsilon N_i^r} \exp(<W_Q^{m(t)}h_i^{(t)},W_{rV1}^{m(t)}h_k^{(t)}>) 
+ 
\]
\begin{equation}
\exp(<W_Q^{m(t)}h_i^{(t)},W_{VS}^{m(t)}S^{(t)}>)
+ \exp(<W_Q^{m(t)}h_i^{(t)},W_{V0}^{m(t)}h_i^{(t)}>),\;
\end{equation}

and the attention coefficients can thus be represented by
 
\[
\alpha_{ijr}^m = \frac{exp(<W_Q^{m(t)}h_i^{(t)},W_{rV1}^{m(t)}h_j^{(t)}>)}{\hat{\sum}^m},
\]

\[
\alpha_{is,r=s}^m = \frac{exp(<W_Q^{m(t)}h_i^{(t)},W_{VS}^{m(t)}S^{(t)}>)}{\hat{\sum}^m},
\]

\[
\alpha_{i0,r=0}^m = \frac{exp(<W_Q^{m(t)}h_i^{(t)},W_{V0}^{m(t)}h_i^{(t)}>)}{\hat{\sum}^m}.
\]

It should be noted that for all relations, we adopt a parameter sharing scheme  \(W_{K*}=W_{V*}\).

\vspace{2mm}

\textbf{Step 3: Star Update}

Following step 1 and step 2, after we got the real node representation at layer \(t+1\), we can now update the star nodes via

\begin{equation}
S^{(t+1)} = \bigparallel_m^{N_{Head}}\sigma(\sum_{j\epsilon N_b} \alpha_{s,j}^m W_{V}^{m(t)} h_j^{(t+1)}+\alpha_{s,s}^m W_{V}^{m(t)} S^t).
\end{equation}

We define

\[\hat{\sum}_s^m = \sum_{k\epsilon N_b} \exp(<W_Q^{m} S^t,W_K^m h_k^{(t+1)}>) + \exp(<W_Q^{m(t)} S^t,W_K^m S^t>)\; ,and\]
\vspace{1mm}

the attention coefficients are thus
\[\alpha_{s,j}^m = \frac{exp(<W_Q^{m}S^{(t)},W_K^{m}h_j^{(t+1)}>)}{\hat{\sum}_s^m},\]
\[\alpha_{s,s}^m = \frac{exp(<W_Q^{m}S^{(t)},W_K^{m}S^{(t)}>)}{\hat{\sum}_s^m}.\]

\section{Multi-Task Setup}

\textbf{Node Classification}

For node classification tasks, following standard graph neural setup described in \cite{velivckovic2017graph,schlichtkrull2018modeling} we take the GraphStar model's final layer output in (2) and apply the standard \(softmax\) activation function. We then minimize the cross-entropy loss conventional in classification model training tasks.

\textbf{Graph Classification}

From the output of (4), in GraphStar, star and this virtual node are connected via other real nodes through attention mechanism, naturally forming hierarchical representation of the graph, meaning \(S^(t)\) is a natural representation of the graph's global state. We use the final layer output of the star for graph classification.

\textbf{Link Predictions}

Here we present the application of GraphStar in single-relation data sets. From (2) We treat the GraphStar model as an encoder for node embedding, and link prediction is a downstream task. A standard setup is to make predictions on subject, relation, object \((s,r,o)\) as described in \cite{schlichtkrull2018modeling}. The decoder is a score function  \(f(s,r,o):\mathbb{R}^d \times\) R \(\times \mathbb{R}^d \rightarrow \mathbb{R}\). The score function for the single-relation experiments is the DistMult \cite{yang2014embedding}: \(f(s,r,o)=e_s^TR_re_o\).

Following  \cite{yang2014embedding, trouillon2016complex,schlichtkrull2018modeling}, we train the model with negative sampling, and for a given positive example, we sample one negative sample through a randomized method to the subject or the object, with cross-entropy loss function.

\section{Experimental Evaluation}

We demonstrate the results for node classification, graph classification and link prediction tasks. We also developed several graph data sets based on standard text classification and sentiment analysis dataset IMDB. 

\subsection{Datasets}

\textbf{Node Classification Tasks}. For node classification, GraphStar is tested on three popular transductive learning benchmarks - Cora, Citeseer and PubMed, and one inductive learning benchmark - PPI.

\textbf{Topic-Specific Sentiment Analysis}. Here, we propose a new method to tackle sentiment analysis based on node classification and use IMDB-binary dataset as an example. This dataset was originally not a graph task; it is usually treated as a natural language processing task (NLP). It is a binary sentiment analysis consisting of 50,000 reviews labeled as either positive or negative. 

Our innovation is to turn the pure NLP task into a graph data task based on document embedding. First, for IMBD, we use a pre-trained large Bert model to document encoding \cite{devlin2018bert}, and we treat every film review as a node in a graph. We then link the nodes (film reviews) which belong to the same topic and create a graph dataset. This approach is highly generalizable to most topic-specific classification tasks. 

\begin{table}[!ht]
\centering
\center{\includegraphics[width=5.5in]
      {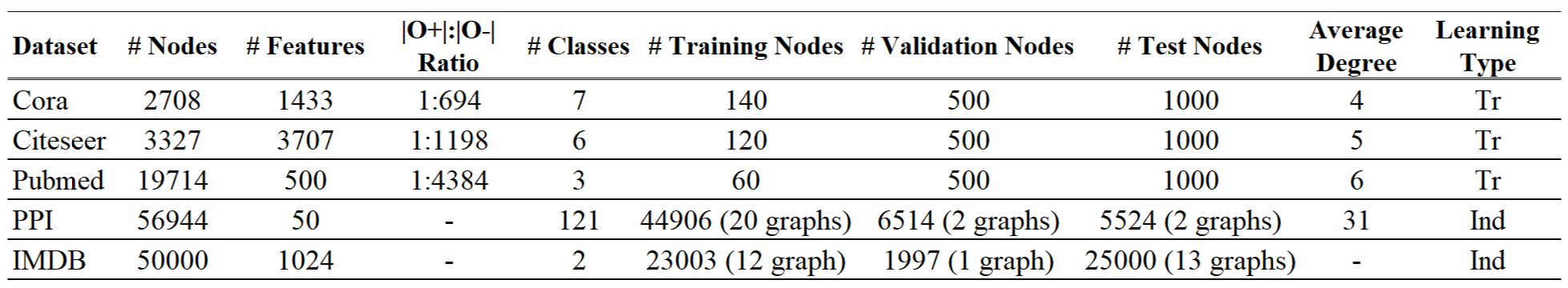}}
\rule{\linewidth}{0cm}

\caption{\label{Summary of Classification Datasets for Node Classification and Link Predictions} Summary of Classification Datasets for Node Classification and Link Predictions.The notation |O+|:|O-| describes the ratio of positive and negative edges. More details regarding these datasets can be found in \cite{sen2008collective, wang2016structural}}
\end{table}

\textbf{Graph Classification task}. For graph classification tasks, we use Enzymes, D\&D, Proteins and Mutag as our evaluation benchmarks. The description of the data is summarized in Table \ref{Table3:Graph Classification Datasets}.

\textbf{Text Classification Tasks}. For text classification, we use R8, R52, MR, 20 News Group and Ohsumed as our benchmarks.
Different from previous text graph-based models\cite{DBLP:journals/corr/abs-1902-07153, yao2018graph}, here we conduct the text classification as graph classification instead of node classifications. In a real life setting, a trained model should be inductive, which can be used to directly make predictions on new texts or documents. In the GraphStar setup, every document corresponds to one graph, and in the graph, every node is a word. If two words appear in the same sliding window, then these two words are linked via an edge. The same words with different context are treated as different nodes. In this way, although the initial word share the same feature, the outputs embedding of the words will be different. For 20 News Group, due to the text length, maximum being over 15000, same words are treated as the same node, and we truncate the text length at 3072. We divide the original training data into 90\% of training data and 10\% of validation data, leaving the test data out of the training cycles. For our experiments, we report the test data results. The summary of the data is in Table \ref{Table2:Graph Text Classification Datasets}.

\begin{table}[!ht]
\centering
\center{\includegraphics[width=3.5in]
      {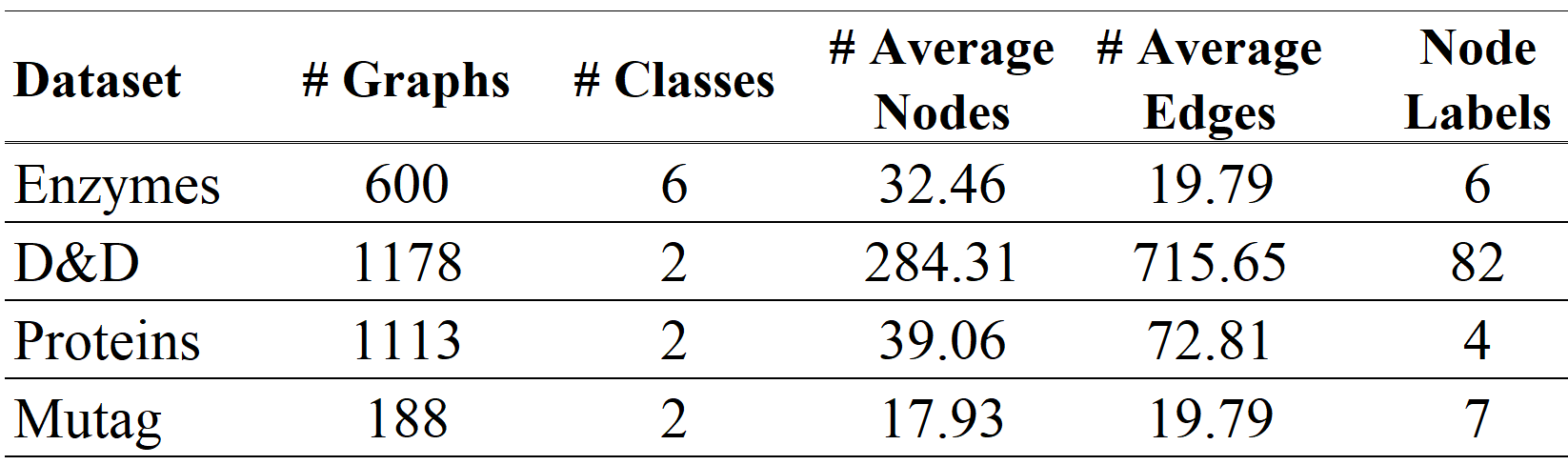}}
\rule{\linewidth}{0cm}
\caption{\label{Table3:Graph Classification Datasets} Graph Classification Datasets}
\end{table}
\vspace*{-\baselineskip}

\begin{table}[!ht]
\centering
\center{\includegraphics[width=5.5in]
      {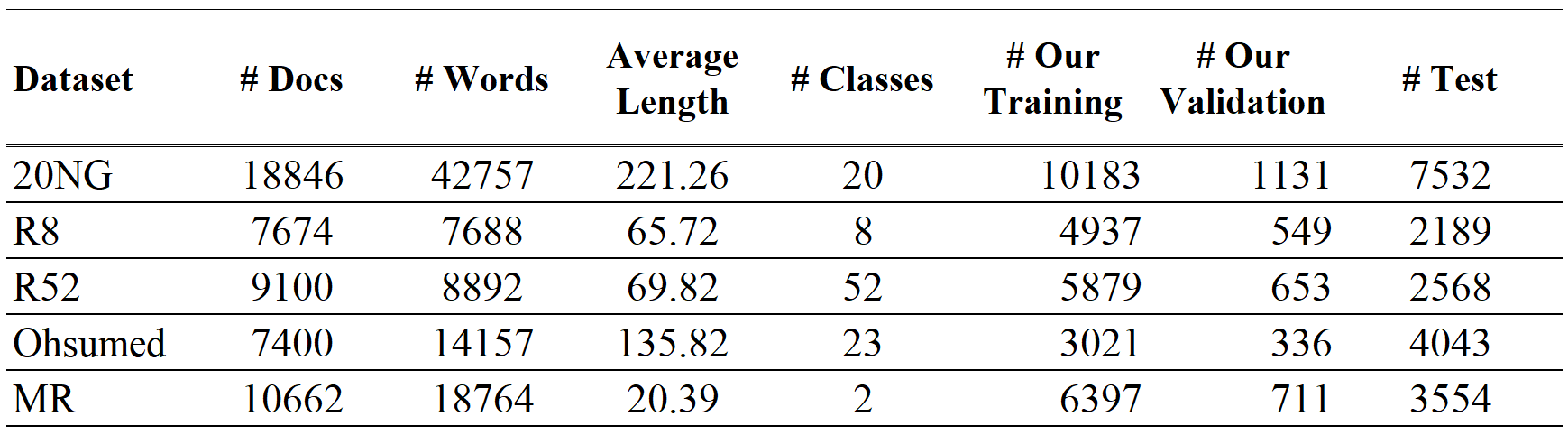}}
\rule{\linewidth}{0cm}
\caption{\label{Table2:Graph Text Classification Datasets}Summary of Text Classification Datasets for Graph Classification}
\end{table}

\textbf{Link Prediction}. For link prediction, following \cite{DBLP:journals/corr/abs-1811-02798}, we use Cora, Citeseer, Pubmed as key benchmarks, and the data description is in Table \ref{Summary of Classification Datasets for Node Classification and Link Predictions}.

\subsection{Results}

The experimental results are summarized in Table \ref{Table1:Result 1}, \ref{Table2:Result IMDB}, \ref{Table3:Graph Classification}, \ref{Table4:Graph Classification Text} and \ref{Table5:Link Prediction} for transductive node classification, inductive node classification, graph classification, graph text classification and link prediction. All reported results are based on 10 runs. We adopt the \textit{Adam} optimizer, and the margins of error are provided in the parenthesis.

\textbf{Node Classification}

For the transductive tasks, all tasks share the same parameters and are not optimized individually for different datasets, and the only variations are the initial learning rates and \(L_2\) regularization settings. We use the attention coefficient dropout rate \(= 0.2\). The number of GraphStar layers \(=2\), the number of attention head per layer \(N_{head}=8\), hidden unit size per head is 16, and the hidden unit's dropout rate is 0.7. For the training, we keep the best model achieving the highest validation accuracy given the patience setting of 50 epochs. The initialization of the star nodes is based on the mean of raw features of the nodes. During training, for Cora, Citeseer and Pubmed, we apply the \(L2\) regularization with \(\rho = 0.002\), \(\rho = 0.004\), and \(\rho = 0.0001\), and the initial learning rates are set to be 0.001, 0.001 and 0.005, respectively. The initial learning rates for PPI and IMDB are 0.0002 and 0.0001, respectively. The \(L2\) setting is \(\rho = 0\) for both PPI and IMDB. For both data sets, the attention coefficient and the hidden unit dropout rates are both 0.2 with 3 layers. The number of hidden units per head is 128 for PPI and 256 for IMDB.

For transductive tasks, the GraphStar classification accuracy is measured against the results in \cite{gao2018large,velivckovic2017graph,DBLP:journals/corr/abs-1902-07153,DBLP:journals/corr/abs-1811-02798,zhang2018gaan}. For inductive task the reported results are in Table \ref{Table1:Result 1}, based on \cite{gao2018large, hamilton2017inductive,velivckovic2017graph,xu2018representation}; the GraphStar model achieves better performance than the current state-of-the-art models for the PPI dataset.

\begin{table}[!ht]
\centering
\center{\includegraphics[width=5.5in]
      {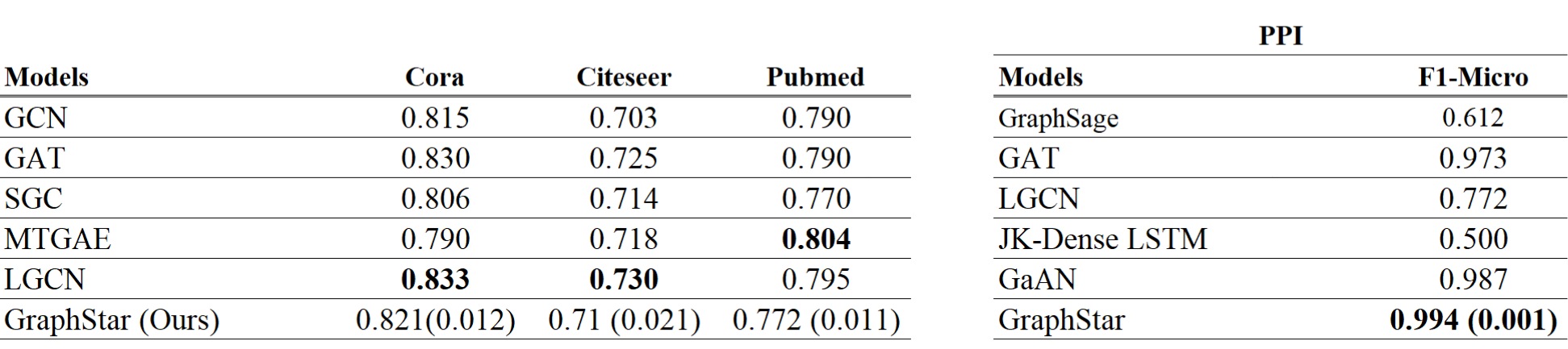}}
\rule{\linewidth}{0cm}
\caption{\label{Table1:Result 1} Results of transductive and inductive node classification experiments. Left: Results of transductive experiments on Cora, Citeseer and Pubmed datasets for multiple landmark model architectures \cite{gao2018large,velivckovic2017graph,DBLP:journals/corr/abs-1902-07153,DBLP:journals/corr/abs-1811-02798,zhang2018gaan}. Right: Results of inductive experiment in terms of micro-average F1 scores on the PPI dataset \cite{gao2018large, hamilton2017inductive,velivckovic2017graph,xu2018representation}}.
\end{table}
\vspace*{-\baselineskip}
\begin{table}[!ht]
\centering
\center{\includegraphics[width=2.2in]
      {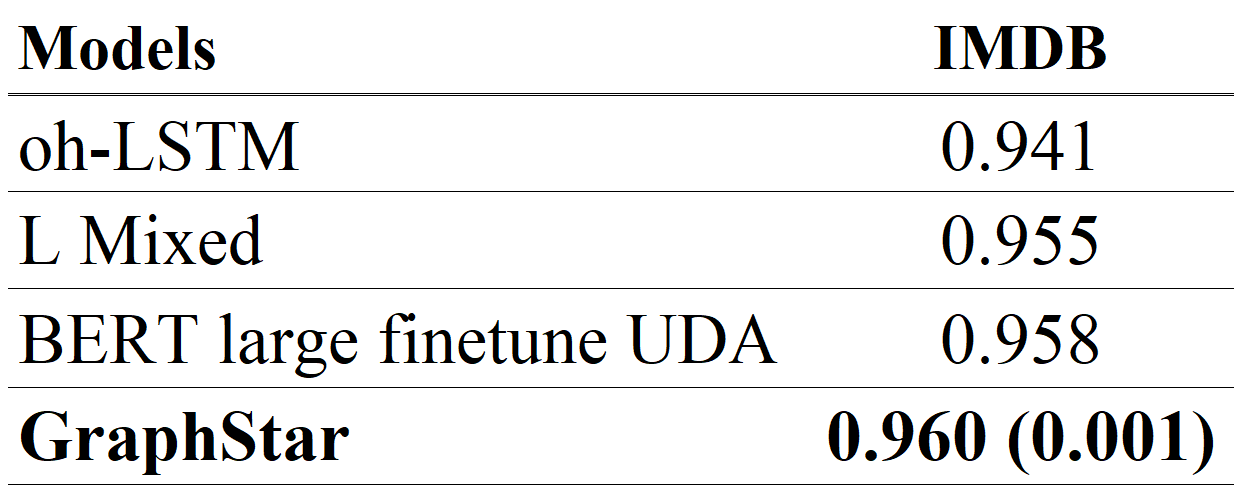}}
\rule{\linewidth}{0cm}
\caption{\label{Table2:Result IMDB} Results of inductive IMDB node classification experiments with reported results based on \cite{johnson2016supervised,sachan2018revisiting,2019arXiv190412848X}}
\end{table}

The IMDB sentiment analysis is reported in Table \ref{Table2:Result IMDB}. The GraphStar is shown to outperform the current state-of-the-art results after only 130 seconds of training time on V100 GPU.

\textbf{Graph Classification / Graph-based Text Classification}. 

For graph classification, we report the current state-of-the-art results in Table \ref{Table3:Graph Classification}. For text classification, the current state-of-the-art results are reported and compared against our model\cite{verma2018graph,ying2018hierarchical,li2019semi} and reported in Table \ref{Table4:Graph Classification Text}.

For Enzymes, Proteins, D\&D and Mutag classification tasks, we use the attention coefficient dropout rate and the hidden unit dropout rate of \(0.2\). The number of GraphStar layers \(=3\), the number of attention head \(N_{head}=4\). The learning rates are 0.0005 for all datasets. The \(L2\) regularization is 0.0001, 0.001, 0.001, 0.0 and the hidden unit sizes per head are 16, 128, 16 and 16 for Enzymes, Proteins, D\&D and Mutag, respectively. For all datasets, we perform 10-fold cross-validation for model evaluation, and the resulting accuracy and margins are based on the 10 runs.

For the text classification tasks, we use the attention coefficient dropout rate \(= 0.3\). The number of GraphStar layers \(=3\), the number of attention head \(N_{head}=4\), hidden unit size per head is 128, and the hidden unit's dropout rate is 0.3. For the training, we keep the modes with the best validation accuracy given the patience setting of 20 epochs. During training, we apply the learning rate of 0.01 and \(L2\) regularization with \(\rho = 0.002\) for all datasets. 

\begin{table}[!ht]
\centering
\center{\includegraphics[width=4in]
      {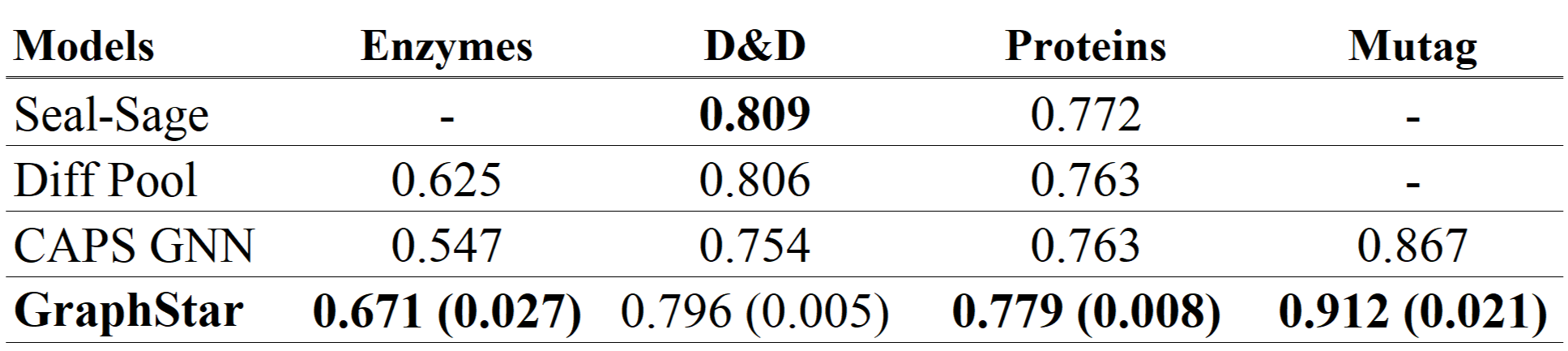}}
\rule{\linewidth}{0cm}
\caption{\label{Table3:Graph Classification} Results of Graph Classification tasks. Left: Science datasets. Right: Text classification datasets.}
\end{table}
\begin{table}[!ht]
\centering
\center{\includegraphics[width=5.5in]
      {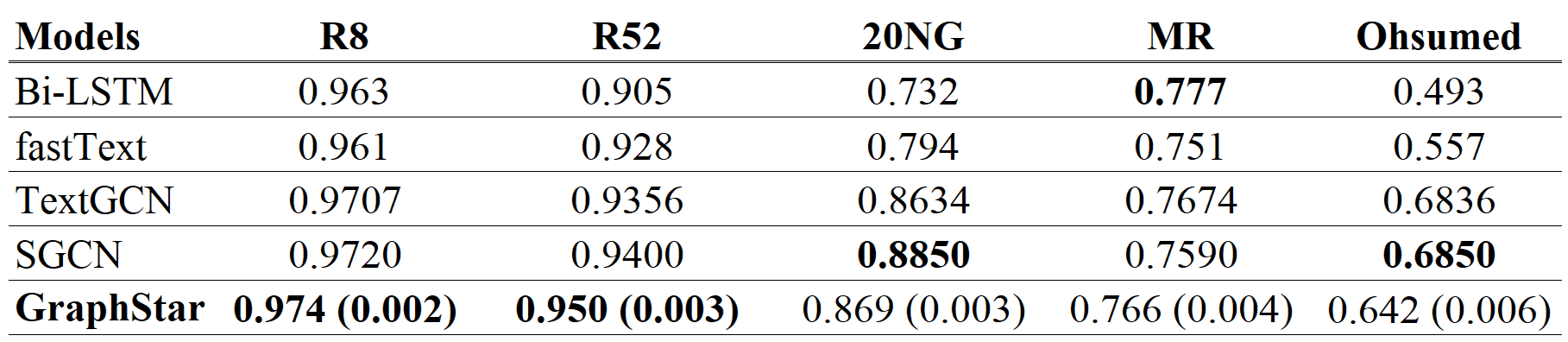}}
\rule{\linewidth}{0cm}
\caption{\label{Table4:Graph Classification Text} Results of graph text classification tasks.}
\end{table}
\vspace*{-\baselineskip}

\textbf{Link Prediction Task}. 

The evaluation metric for link prediction task is the average of the AUC and AP scores. The number of attention head \(N_{head}=4\), the number of layer is 3, initial learning rate is 0.0002, and the \(L2\) regularization setting is 0.0005. The same setting is applied to all datasets for link prediction except the hidden size per layer and dropout. The attention coefficient dropout is 0 for all datasets, and the hidden layer dropout rate is 0.2 for Cora and 0 for Citeseer and Pubmed. The hidden unit size per head is 512 for Cora and Citeseer, and 128 for Pubmed. For the training, we keep the model achieving the highest average of the AUC and AP scores on the validation set, given the patience setting of 300 epochs. 

For link prediction, the \cite{kipf2016semi, DBLP:journals/corr/abs-1811-02798} state-of-the-art results are reported. Link prediction performance is reported as the combined average of AUC and AP scores.

\begin{table}[!ht]
\centering
\center{\includegraphics[width=3.2in]
      {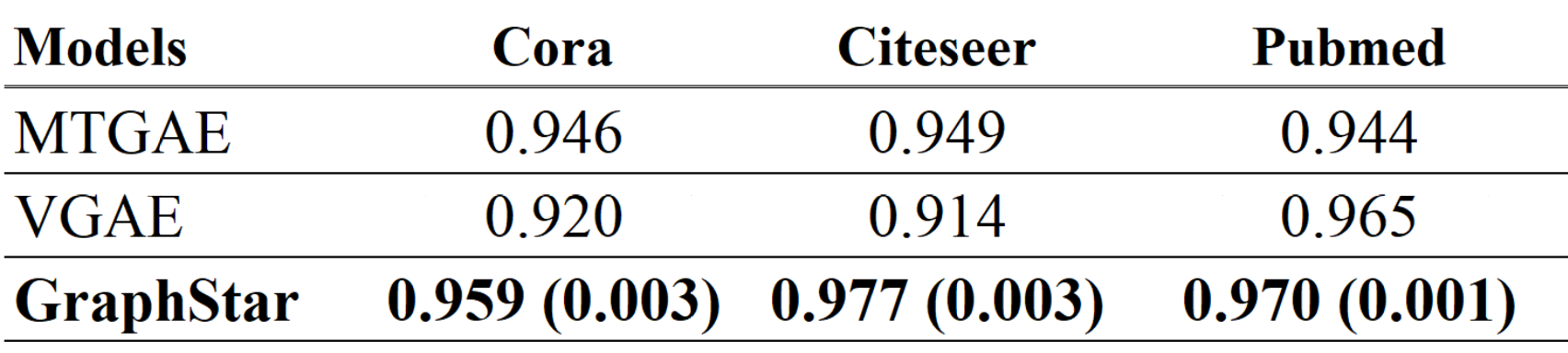}}
\rule{\linewidth}{0cm}
\caption{\label{Table5:Link Prediction} Results of link prediction experiments.}
\end{table}

\subsection{Analysis of Results}
In this section, we study the inductive node classification training size effect and the stability of graph classification. For PPI dataset, we gradually reduce the original training set graph size and give them to validation and test sets. The result is reported in Table \ref{Table:Training effectiveness for gradual training set reduction}. 

The result indicates that for PPI, GraphStar is particular effective such that, even if we reduce the training size drastically, the training accuracy can still achieve reasonable performance. We also observe the same effect in the IMDB dataset such that, by reducing the original training size of 25000 to 1997 and treating the 23003 as extra validation set, the model still maintains the accuracy of 94.0\% on the test set. 

\begin{table}[!ht]
\centering
\center{\includegraphics[width=3in]
      {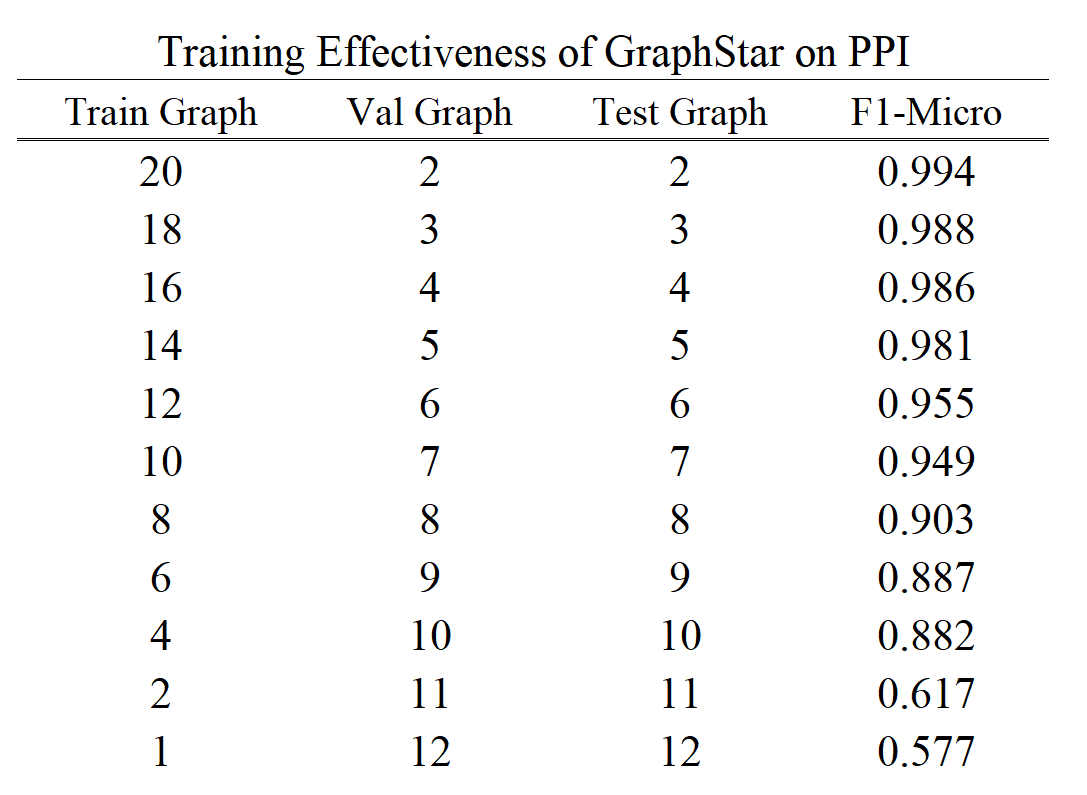}}
\rule{\linewidth}{0cm}
\caption{\label{Table:Training effectiveness for gradual training set reduction} Result of training effectiveness of GraphStar and training size reduction on PPI dataset}
\end{table}

\begin{figure}[!ht]
\centering
\center{\includegraphics[width=5.5in]
      {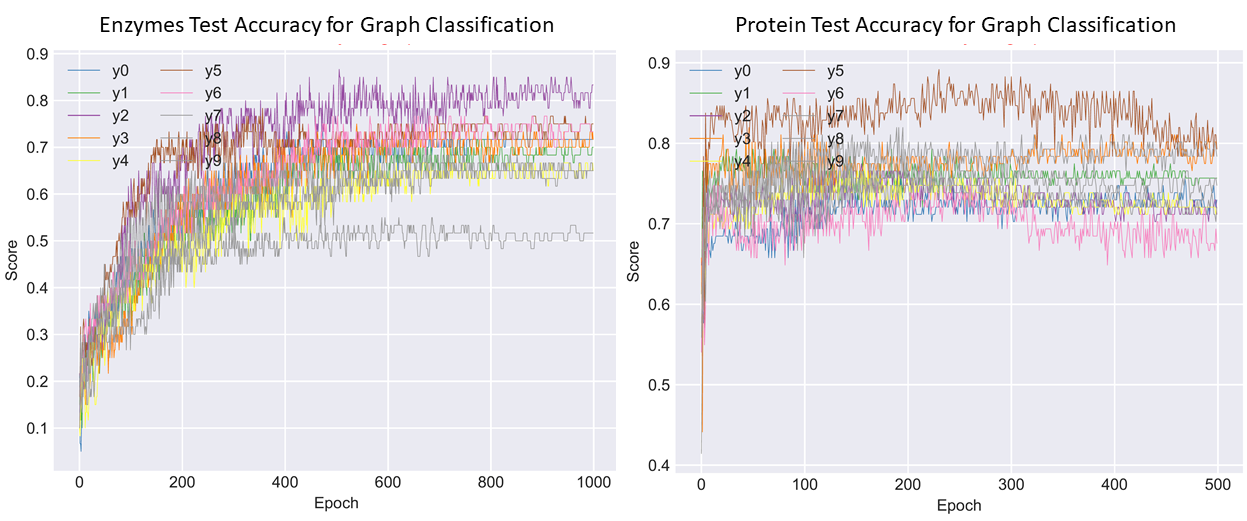}}
\rule{\linewidth}{0cm}
\caption{\label{Figure:Analysis for test accuracy trends in 10-fold cross validation for Enzymes and Protein} Stability for test accuracy in 10-fold cross-validation training in graph classification. Left: Result for Enzymes dataset. Right: Result for Protein dataset}
\end{figure}

Additionally, we also study the stability of 10-fold cross validation process for graph classification and present it in Figure \ref{Figure:Analysis for test accuracy trends in 10-fold cross validation for Enzymes and Protein}. The process for Enzyme and Protein data sets are presented here, and as the graph indicates, the experiments are conducted on the 10-fold validation across epochs. All reported accuracy are based on experiment setups described above. The figure indicates that for graph classification, the test results can have high fluctuation for those datasets, and our best results, compared with the current state-of-the-art models, have considerably higher accuracy.

\section{Conclusion}
We have presented graph star net (GraphStar) which can be used in three general graph tasks (node classification, graph classification and link prediction). GraphStar, without incurring heavy computation costs, is able to map the global state effectively. Lastly, we also propose novel approaches to conduct topic-specific sentiment analysis and text classification with graph models. GraphStar has achieved state-of-the-art performance across all tasks.

\newpage 
\bibliographystyle{plain}
\bibliography{GraphStarNet}

\begin{thebibliography}{10}

\bibitem{devlin2018bert}
Jacob Devlin, Ming-Wei Chang, Kenton Lee, and Kristina Toutanova.
\newblock Bert: Pre-training of deep bidirectional transformers for language
  understanding.
\newblock {\em arXiv preprint arXiv:1810.04805}, 2018.

\bibitem{gao2018large}
Hongyang Gao, Zhengyang Wang, and Shuiwang Ji.
\newblock Large-scale learnable graph convolutional networks.
\newblock In {\em Proceedings of the 24th ACM SIGKDD International Conference
  on Knowledge Discovery \& Data Mining}, pages 1416--1424. ACM, 2018.

\bibitem{gilmer2017neural}
Justin Gilmer, Samuel~S Schoenholz, Patrick~F Riley, Oriol Vinyals, and
  George~E Dahl.
\newblock Neural message passing for quantum chemistry.
\newblock In {\em Proceedings of the 34th International Conference on Machine
  Learning-Volume 70}, pages 1263--1272. JMLR. org, 2017.

\bibitem{guo2019star}
Qipeng Guo, Xipeng Qiu, Pengfei Liu, Yunfan Shao, Xiangyang Xue, and Zheng
  Zhang.
\newblock Star-transformer.
\newblock {\em arXiv preprint arXiv:1902.09113}, 2019.

\bibitem{hamilton2017inductive}
Will Hamilton, Zhitao Ying, and Jure Leskovec.
\newblock Inductive representation learning on large graphs.
\newblock In {\em Advances in Neural Information Processing Systems}, pages
  1024--1034, 2017.

\bibitem{johnson2016supervised}
Rie Johnson and Tong Zhang.
\newblock Supervised and semi-supervised text categorization using lstm for
  region embeddings.
\newblock {\em arXiv preprint arXiv:1602.02373}, 2016.

\bibitem{kipf2016semi}
Thomas~N Kipf and Max Welling.
\newblock Semi-supervised classification with graph convolutional networks.
\newblock {\em arXiv preprint arXiv:1609.02907}, 2016.

\bibitem{li2019semi}
Jia Li, Yu~Rong, Hong Cheng, Helen Meng, Wenbing Huang, and Junzhou Huang.
\newblock Semi-supervised graph classification: A hierarchical graph
  perspective.
\newblock 2019.

\bibitem{li2015gated}
Yujia Li, Daniel Tarlow, Marc Brockschmidt, and Richard Zemel.
\newblock Gated graph sequence neural networks.
\newblock {\em arXiv preprint arXiv:1511.05493}, 2015.

\bibitem{monti2017geometric}
Federico Monti, Davide Boscaini, Jonathan Masci, Emanuele Rodola, Jan Svoboda,
  and Michael~M Bronstein.
\newblock Geometric deep learning on graphs and manifolds using mixture model
  cnns.
\newblock In {\em Proceedings of the IEEE Conference on Computer Vision and
  Pattern Recognition}, pages 5115--5124, 2017.

\bibitem{perozzi2014deepwalk}
Bryan Perozzi, Rami Al-Rfou, and Steven Skiena.
\newblock Deepwalk: Online learning of social representations.
\newblock In {\em Proceedings of the 20th ACM SIGKDD international conference
  on Knowledge discovery and data mining}, pages 701--710. ACM, 2014.

\bibitem{sachan2018revisiting}
Devendra~Singh Sachan, Manzil Zaheer, and Ruslan Salakhutdinov.
\newblock Revisiting lstm networks for semi-supervised text classification via
  mixed objective function.
\newblock 2018.

\bibitem{schlichtkrull2018modeling}
Michael Schlichtkrull, Thomas~N Kipf, Peter Bloem, Rianne Van Den~Berg, Ivan
  Titov, and Max Welling.
\newblock Modeling relational data with graph convolutional networks.
\newblock In {\em European Semantic Web Conference}, pages 593--607. Springer,
  2018.

\bibitem{sen2008collective}
Prithviraj Sen, Galileo Namata, Mustafa Bilgic, Lise Getoor, Brian Galligher,
  and Tina Eliassi-Rad.
\newblock Collective classification in network data.
\newblock {\em AI magazine}, 29(3):93--93, 2008.

\bibitem{tang2015line}
Jian Tang, Meng Qu, Mingzhe Wang, Ming Zhang, Jun Yan, and Qiaozhu Mei.
\newblock Line: Large-scale information network embedding.
\newblock In {\em Proceedings of the 24th international conference on world
  wide web}, pages 1067--1077. International World Wide Web Conferences
  Steering Committee, 2015.

\bibitem{DBLP:journals/corr/abs-1811-02798}
Phi~Vu Tran.
\newblock Multi-task graph autoencoders.
\newblock {\em CoRR}, abs/1811.02798, 2018.

\bibitem{trouillon2016complex}
Th{\'e}o Trouillon, Johannes Welbl, Sebastian Riedel, {\'E}ric Gaussier, and
  Guillaume Bouchard.
\newblock Complex embeddings for simple link prediction.
\newblock In {\em International Conference on Machine Learning}, pages
  2071--2080, 2016.

\bibitem{vaswani2017attention}
Ashish Vaswani, Noam Shazeer, Niki Parmar, Jakob Uszkoreit, Llion Jones,
  Aidan~N Gomez, {\L}ukasz Kaiser, and Illia Polosukhin.
\newblock Attention is all you need.
\newblock In {\em Advances in neural information processing systems}, pages
  5998--6008, 2017.

\bibitem{velivckovic2017graph}
Petar Veli{\v{c}}kovi{\'c}, Guillem Cucurull, Arantxa Casanova, Adriana Romero,
  Pietro Lio, and Yoshua Bengio.
\newblock Graph attention networks.
\newblock {\em arXiv preprint arXiv:1710.10903}, 2017.

\bibitem{verma2018graph}
Saurabh Verma and Zhi-Li Zhang.
\newblock Graph capsule convolutional neural networks.
\newblock {\em arXiv preprint arXiv:1805.08090}, 2018.

\bibitem{wang2016structural}
Daixin Wang, Peng Cui, and Wenwu Zhu.
\newblock Structural deep network embedding.
\newblock In {\em Proceedings of the 22nd ACM SIGKDD international conference
  on Knowledge discovery and data mining}, pages 1225--1234. ACM, 2016.

\bibitem{wang2018non}
Xiaolong Wang, Ross Girshick, Abhinav Gupta, and Kaiming He.
\newblock Non-local neural networks.
\newblock In {\em Proceedings of the IEEE Conference on Computer Vision and
  Pattern Recognition}, pages 7794--7803, 2018.

\bibitem{DBLP:journals/corr/abs-1902-07153}
Felix Wu, Tianyi Zhang, Amauri H.~Souza Jr., Christopher Fifty, Tao Yu, and
  Kilian~Q. Weinberger.
\newblock Simplifying graph convolutional networks.
\newblock {\em CoRR}, abs/1902.07153, 2019.

\bibitem{2019arXiv190412848X}
Qizhe {Xie}, Zihang {Dai}, Eduard {Hovy}, Minh-Thang {Luong}, and Quoc~V. {Le}.
\newblock {Unsupervised Data Augmentation}.
\newblock {\em arXiv e-prints}, page arXiv:1904.12848, Apr 2019.

\bibitem{xu2018representation}
Keyulu Xu, Chengtao Li, Yonglong Tian, Tomohiro Sonobe, Ken-ichi Kawarabayashi,
  and Stefanie Jegelka.
\newblock Representation learning on graphs with jumping knowledge networks.
\newblock {\em arXiv preprint arXiv:1806.03536}, 2018.

\bibitem{yang2014embedding}
Bishan Yang, Wen-tau Yih, Xiaodong He, Jianfeng Gao, and Li~Deng.
\newblock Embedding entities and relations for learning and inference in
  knowledge bases.
\newblock {\em arXiv preprint arXiv:1412.6575}, 2014.

\bibitem{yao2018graph}
Liang Yao, Chengsheng Mao, and Yuan Luo.
\newblock Graph convolutional networks for text classification.
\newblock {\em arXiv preprint arXiv:1809.05679}, 2018.

\bibitem{ying2018hierarchical}
Zhitao Ying, Jiaxuan You, Christopher Morris, Xiang Ren, Will Hamilton, and
  Jure Leskovec.
\newblock Hierarchical graph representation learning with differentiable
  pooling.
\newblock In {\em Advances in Neural Information Processing Systems}, pages
  4800--4810, 2018.

\bibitem{zhang2018gaan}
Jiani Zhang, Xingjian Shi, Junyuan Xie, Hao Ma, Irwin King, and Dit-Yan Yeung.
\newblock Gaan: Gated attention networks for learning on large and
  spatiotemporal graphs.
\newblock {\em arXiv preprint arXiv:1803.07294}, 2018.

\bibitem{zhou2018graph}
Jie Zhou, Ganqu Cui, Zhengyan Zhang, Cheng Yang, Zhiyuan Liu, and Maosong Sun.
\newblock Graph neural networks: A review of methods and applications.
\newblock {\em arXiv preprint arXiv:1812.08434}, 2018.

\end{thebibliography}
\end{document}